\documentclass[11pt]{article}
\usepackage{graphicx,floatflt,amssymb,axodraw}
\def\pmb#1{\setbox0=\hbox{#1}
    \kern-.015em\copy0\kern-\wd0
    \kern.03em\copy0\kern-\wd0
    \kern-.015em\raise.0233em\box0}

\newcommand{\nn}{\nonumber}
\def\lsim{\mathrel{\vcenter{\hbox{$<$}\nointerlineskip\hbox{$\sim$}}}}

\renewcommand{\thefootnote}{\fnsymbol{footnote}}
\newcommand{\be}[1]{\begin{equation} \label{(#1)}}
\newcommand{\ee}{\end{equation}}
\newcommand{\ba}[1]{\begin{eqnarray} \label{(#1)}}
\newcommand{\ea}{\end{eqnarray}}
\newcommand{\rf}[1]{(\ref{(#1)})}
\def\lsim{\mathrel{\vcenter{\hbox{$<$}\nointerlineskip\hbox{$\sim$}}}}

\def \Rpv{$R_{P} \hspace{-0.9em}/\;\:\hspace{0em}$}

\def\nonu{neutrinoless double-$\beta$ decay}

\begin{document}
\pagestyle{empty}
\begin{flushright}
\texttt{hep-ph/0302191}\\
WUE-ITP-2003-002\\
SINP/TNP/03-06\\
\end{flushright}

\vskip 100pt

\begin{center}
{\Large \bf Particle Physics Implications of the WMAP Neutrino Mass Bound} \\
\vspace*{1cm}
\renewcommand{\thefootnote}{\fnsymbol{footnote}}
{\large\sf Gautam Bhattacharyya $^{1 \!\!}$
\footnote{E-mail address: gb@theory.saha.ernet.in}},
{\large\sf Heinrich~P\"as $^{2,\!\!}$
\footnote{E-mail address: paes@physik.uni-wuerzburg.de}},
{\large\sf Liguo~Song $^{3,\!\!}$
\footnote{E-mail address: liguo.song@vanderbilt.edu}}, 
{\large\sf Thomas~J.~Weiler $^{3,\!\!}$
\footnote{E-mail address: tom.weiler@vanderbilt.edu}}
\vskip 10pt
$^{1)}${\small \it Saha Institute of Nuclear Physics, 1/AF Bidhan Nagar,
 Kolkata 700064, India}\\
$^{2)}${\small \it Institut f\"ur Theoretische Physik und Astrophysik,
Universit\"at W\"urzburg, D-97074 W\"urzburg, Germany }\\
$^{3)}${\small \it Department of Physics and Astronomy, Vanderbilt University,
Nashville, TN 37235, USA}\\
\normalsize
\end{center}

\begin{abstract}
The recently published cosmological bound on the absolute neutrino
masses obtained from the Wilkinson Microwave Anisotropy Probe (WMAP)
data has important consequences for neutrino experiments and models.
Taken at face value, the new bound excludes the determination of the
absolute neutrino mass in the KATRIN experiment and disfavors a neutrino
oscillation interpretation of the LSND experiment. Combined with the
KamLAND and Super-K data, the WMAP bound defines an accessible range for
the neutrinoless double beta decay amplitude.  The bound also impacts
the Z-burst annihilation mechanism for resonant generation of
extreme-energy cosmic rays on the cosmic neutrino background in two
ways: it constrains the local over-density of neutrino dark matter which
is not helpful, but it also limits the resonant energy to a favorable
range.  In R-parity violating SUSY models, neutrino masses are generated
by trilinear and bilinear lepton number violating couplings. The WMAP
result improves the constraints on these couplings 
by an order of magnitude.

\vskip 5pt \noindent
\texttt{PACS Nos: 12.60.Jv, 14:60.Pq, 14.80.Ly, 23:40.Bw, 96.40.Tv,
98.80.Es}
 \\
\end{abstract}
\vskip 20pt

\newpage

\pagestyle{plain}
\setcounter{page}{1}

\renewcommand{\thesection}{\Roman{section}}
\setcounter{footnote}{0}
\renewcommand{\thefootnote}{\arabic{footnote}}

\section{Introduction}
With the recently published first data of the Wilkinson Microwave Anisotropy
Probe (WMAP) \cite{Spergel:2003cb} on the cosmic microwave background
(CMB) anisotropies the age of precision cosmology has arrived.  A flat,
vacuum-energy dominated cold dark matter
($\Lambda$CDM) universe seeded by nearly
scale-invariant Gaussian primordial fluctuations appears to be firmly
established as the standard cosmology.  Moreover, when combined with
additional CMB data-sets (CBI, ACBAR) \cite{cbi-acbar} and observations
of large scale structure from the 2dF Galaxy Redshift Survey (2dFGRS)
\cite{2dF} to lift degeneracies, the WMAP data offers the potential of
testing various extensions and sub-dominant components in the
$\Lambda$CDM model, such
as small non-flatness, quintessence, possible tensor-gravitational wave
modes, and a massive cosmic neutrino background (C$\nu$B).
Investigation of the latter has most important consequences for
terrestrial physics experiments exploring the neutrino sector.

The power spectrum of early-Universe density perturbations
is processed by gravitational instabilities.
However, 
the free-streaming relativistic 
neutrinos suppress the growth of fluctuations
on scales below the horizon 
(approximately the Hubble size $c/H(z)$) 
until they become nonrelativistic at 
$z\sim m_j/3T_0 \sim 1000\,(m_j/{\rm eV})$.
When the amplitude of fluctuations is normalized to the WMAP data, the 
amplitude of fluctuations in the 2dFGRS places significant limits on the 
contribution of neutrinos to the energy density of the universe,
\be{}
\Omega_{\nu} h^2 = \frac{\sum_i m_i}{93.5~{\rm eV}} < 0.0076
\quad (95\%\,{\rm C.L.}),
\ee
which translates into 
\be{mapbound}
\sum_i m_i < 0.71 ~{\rm eV} \quad (95\%\,{\rm C.L.}).   
\ee

The new mass bound \rf{mapbound} impacts most directly four-neutrino 
mass models constructed to accommodate the LSND evidence for 
oscillation.  Such models require the heaviest neutrino mass to be 
$\sim 1$~eV, and so at face value are disfavored by the new result 
\cite{fournu,giunti}.
However, there are several loopholes in the argument against an $\sim 1$~eV
sterile neutrino.  If there is only one isolated ``heavy'' sterile 
as in the 3+1 model, 
then the WMAP/2dF data at face value allow the  
$\Delta m^2_{\rm LSND}$ region up to 0.5~eV$^2$,
whereas relaxing the WMAP/2dF bound from 0.71~eV to 1~eV allows 
virtually the entire LSND region to co-exist.
In a 2+2 model, there are two heavy mass eigenstates,
and the WMAP/2dF data at face value limit $\Delta m^2_{\rm LSND}$ 
to 0.1~eV$^2$.  
Still another possibility, not yet explored to the best of our knowledge,
might be to model the heavier neutrinos 
as decaying to light flavors plus a light boson,
with a lifetime much less than the age of the Universe at 
structure formation.
In such a model, the decay products would be 
free-streaming particles with masses well
below the WMAP bound.
Relevant to this discussion is the limit from 
Big Bang Nucleosynthesis (BBN) \cite{abazajian,BBN},
that neutrinos beyond the three active could not have been in  
thermal equilibrium already at the BBN temperature $\sim$MeV, long before 
the epoch of structure formation.
So the more serious constraint for the sterile neutrino is the BBN limit.  
Overcoming this BBN limit automatically immunizes the sterile against the
WMAP/2dF bound \cite{DiBariAnd},
since the depopulated states at BBN are not populated at a later time.
One way to evade thermalization at the BBN epoch is 
via a tiny lepton asymmetry \cite{lepasym}.
There are several other ways, conveniently summarized in \cite{abazajian}.
In conclusion, MiniBooNE is still required to settle the fate of the sterile
neutrino \cite{PSW}.

From here on we focus on the consequences 
of the new WMAP bound for three-neutrino models.  
It was previously noted \cite{PW}
that there are potentially four independent 
approaches for 
measuring the absolute neutrino mass.
These are large-scale structure studies measuring the total mass 
in the C$\nu$B (as reported by WMAP), 
the Z-burst method measuring individual masses in the C$\nu$B, 
and the terrestrial measurements of the tritium end point spectrum
and neutrinoless double beta decay rate.
Of course, the results of these approaches are correlated in the sense that 
the experiments all attempt to determine the same neutrino masses.
We will examine the impact of the new WMAP bound on the future 
of the other three approaches.

\begin{figure}
\centerline{\resizebox{12cm}{7cm}{\includegraphics{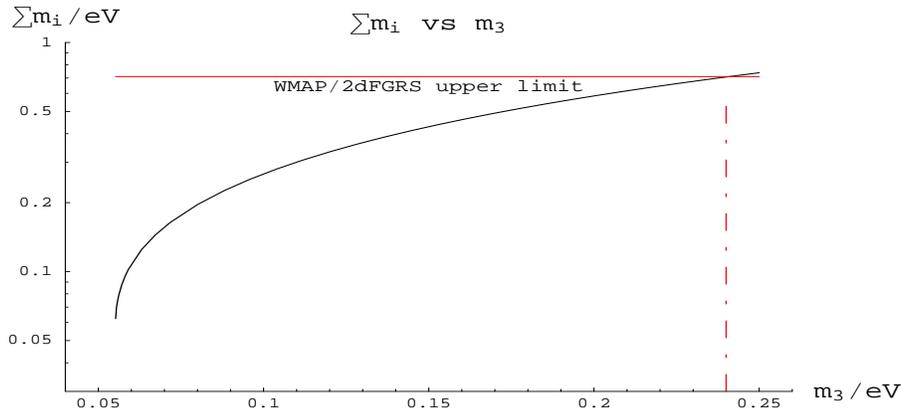}}}
 \caption{\small Implications of the WMAP neutrino mass bound for the mass
of the heaviest neutrino $m_3$. Here we take the best-fit value for 
$\Delta m^2_{\rm atm}=3\times 10^{-3}\,{\rm eV}^2$;
$\Delta m^2_{\rm sun}$ is too small to be relevant.
}
 \label{fig:mapbound}
\end{figure}

Neutrino oscillation studies have established three important facts
of relevance here. The first is that the two mass-squared differences 
are small compared to the WMAP limit.  Thus, when the WMAP limit 
is saturated, the three neutrinos are nearly degenerate in mass,
and we have 
\be{mapmass}
m_i  < 0.24 ~{\rm eV} \quad (95\%\,{\rm C.L.})  
\ee
for each neutrino mass.
The second important fact is that oscillation studies provide a 
{\sl lower} bound on the heaviest neutrino mass, given by the minimum  
$\sqrt{\Delta m^2_{\rm atm}}\sim 0.03$~eV.
Thus, we may write
\be{m3limits}
0.03 ~{\rm eV} \le m_3 \le 0.24 ~{\rm eV} \quad (95\%\,{\rm C.L.})\,,
\ee
which shows the remarkable fact that knowledge of the 
heaviest neutrino mass (which we shall always denote by $m_3$) 
is now known to an order of magnitude!
A plot of the total neutrino mass versus $m_3$
is shown in Fig.~\ref{fig:mapbound}.
The relation is linear, $\sum_i m_i=3\,m_3$,
except very near the smallest allowed $m_3$, 
$\sim\sqrt{\Delta m^2_{\rm atm}}$.
The third important fact is that the three angles parameterizing the unitary 
flavor-mass mixing-matrix, $U_{\alpha i}$, are well known.
The one CP-violating Dirac phase and two CP-violating Majorana phases
are not known.
The angles and phases will be important when we look at neutrinoless
double beta decay. 

Absolute neutrino mass bounds also constrain all entries in the neutrino
mass matrix in flavor space due to unitarity. This results in bounds on
couplings in theories with lepton number violation \cite{gautam}.  As an
example, we derive bounds on parameters of the R-parity violating (\Rpv)
SUSY model, improving them by one order of magnitude over the existing
values.

\section{Tritium beta decay}
The mass to be inferred from $\beta$-decay is
$m^2_{\nu_e}\equiv \sum_j\,|U_{ej}|^2\,m^2_j$.
The KATRIN project \cite{katrin}
plans to start taking data in 2007.
The sensitivity aim after three years of 
measurement is 0.08~eV$^2$ at 1 $\sigma$ accuracy.
This may be improved to 0.05-0.06~eV$^2$, when 
optimizing the data point distribution and resolution,
which implies 
a final sensitivity of $m_{\nu_e}$ to be 0.4~eV at 3 $\sigma$.
Thus, the reach of this experiment includes only the 
nearly mass-degenerate neutrino case, 
for which unitarity allows one to write
$m_{\nu_e}= m_3$.

Comparing the KATRIN reach to the WMAP limit in Eq.\ \rf{mapmass},
one comes to the unfortunate conclusion that a positive signal is unlikely.

\begin{figure}
\centerline{\resizebox{12cm}{7cm}{\includegraphics{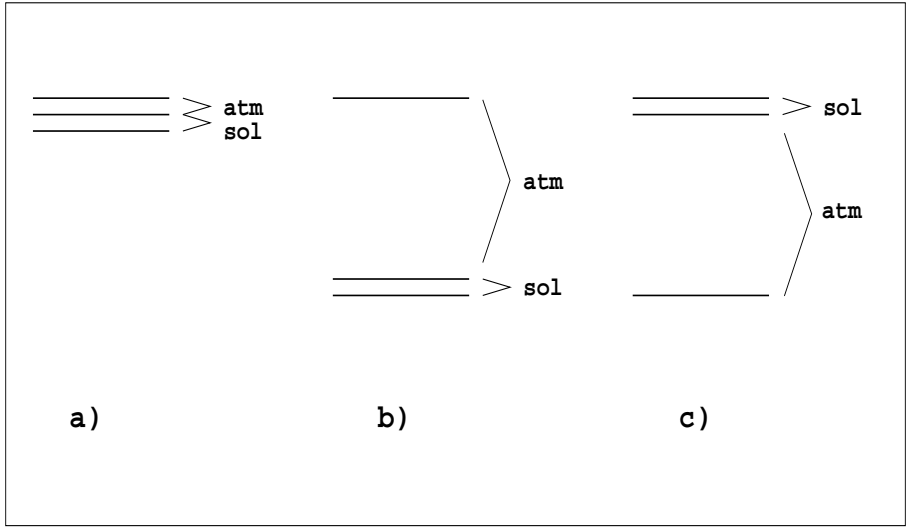}}}
 \caption{\small Neutrino mass spectra for the three neutrino case:
a) degenerate, b) hierarchical and c) inverse hierarchical spectrum.
}
 \label{fig:hierarchy}
\end{figure}


\section{Neutrinoless double beta decay}
The mass inferred in \nonu\ is
\be{nonumass}
m_{ee}= |\sum_i U_{ei}^2 m_i|\,.
\ee
Here one needs the neutrino mixing parameters explicitly.
The most recent analysis of atmospheric neutrino 
data \cite{Fornengo:2001pm} 
yields
\be{}
10^{-3}~{\rm eV^2} < \Delta m_{\rm atm}^2 < 5 \cdot 10^{-3}~{\rm eV^2} 
\ee
and
\be{}
\sin^2 2 \theta_{\rm atm} > 0.8\\.
\ee
On the other hand, 
a recent evaluation of solar neutrino data including the
KamLAND reactor experiment \cite{solar} 
inferred
\be{}
5 \cdot 10^{-5}~{\rm eV^2}< \Delta m_{\rm sun}^2 < 1.1 
\cdot 10^{-4}~{\rm eV^2}.
\ee
and 
\be{}
0.3 < \tan^2 \theta_{\rm sun} < 0.8 \,.
\ee
Thus, the LMA solar solution is confirmed.
The neutrino mixing matrix is seen to be ``bi-large.''
It is also known that $|U_{e3}|^2 \approx 0$,
which means that 
the third mixing angle is negligibly small \cite{chooz}.

The cases of  degenerate, hierarchical, and inverse hierarchical
neutrinos (see Fig. \ref{fig:hierarchy}) must be considered separately
(for a detailed discussion, see e.g. \cite{kps}).
The WMAP limit is sufficiently large that it impacts only
the case of degenerate neutrinos.

\begin{itemize}
\item Degenerate Neutrinos: $m_1 \simeq m_2 \simeq m_3$.
With $|U_{e3}|^2 \approx 0$, one has a mass proportional to 
$|U_{e1}^2 +U_{e2}^2|=|\cos^2\theta+e^{2i\delta}\sin^2\theta|$,
which, upon extremizing the unknown phase, leads to 
\be{}
\cos 2 \theta_{\rm sun} m_3 < m_{ee} < m_3 \,.
\ee
Inputting the new WMAP bound, and the solar angle, one gets 
\be{degenmass}
0.1~m_3 <  m_{ee} < 0.24~{\rm eV} \,.
\ee

\item Hierarchical neutrinos: $m_1 \ll m_2 \ll m_3$
and $\Delta m^2_{\rm sun}= \Delta m^2_{12}$\,.
Here a lower limit is obtained by taking $|U_{e3}|^2=0$
and $m_1=0$.  The result is 
\be{}
m_{ee} > \sqrt{\Delta m^2_{\rm sun}} \sin^2 \theta_{\rm sun}= 
2 \cdot 10^{-3}~{\rm eV},
\ee
and 
$m_{ee}\ll m_3\sim \sqrt{\Delta m^2_{\rm atm}}\lsim 0.07$~eV.

\item Inverse hierarchical neutrinos: $m_1 \ll m_2 \simeq m_3$
and $\Delta m^2_{\rm sun}= \Delta m^2_{23}$. 
The situation is analogous to the degenerate case, 
but with the scale of $m_3$
fixed by the atmospheric neutrino evidence, rather than the WMAP result.
One gets 
\be{}
\cos 2 \theta_{\rm sun} \sqrt{\Delta m^2_{\rm atm}} < m_{ee} < 
\sqrt{\Delta m^2_{\rm atm}}\,,
\ee
i.e.
\be{}
3 \cdot 10^{-3}~{\rm eV} <  m_{ee} < 0.07~{\rm eV} \,.
\ee

\end{itemize}

In summary, neglecting unnatural cancellations due to a conspiracy of 
$\delta$, $m_1$ and mixing angles, the predicted range of
$m_{ee}$ is given by
\be{}
2 \cdot 10^{-3}~{\rm eV} <  m_{ee} < 0.24 ~{\rm eV}.
\ee
Fortunately, the whole region can be covered by the most ambitious
double beta decay proposals \cite{GENIUS} (for an overview
of the experimental status see \cite{elliot}).
The lower limit is not impacted by the WMAP result,
whereas the upper limit comes directly from the WMAP data.
The central value of the 
recent discovery claim of the Heidelberg-Moscow experiment
\cite{evi}, $m_{ee}=0.39^{+0.45}_{-0.34}$~eV, exceeds 
the WMAP bound,
but the reported lower range does not 
(this fact has been pointed out already in 
ref. \cite{fournu}).
We point out, though, that double beta decay mechanisms other than the 
standard neutrino 
mass mechanism are not affected by this bound.
A particular interesting possibility to accommodate the Heidelberg-Moscow 
result involves singlet neutrinos 
propagating in large extra dimensions in which case a mechanism 
decorrelating the neutrino mass eigenstates from the double beta decay
amplitude is operative \cite{extra}. 
Exchange of superpartners in R-parity violating SUSY, leptoquarks, or 
right-handed W bosons constitute other possibilities to account for a
sizable neutrinoless double beta decay signal 
(for a review see \cite{znbbrev}).

\section{The Z-burst model for EECR's}

\begin{figure}
\centerline{\resizebox{12cm}{7cm}{\includegraphics{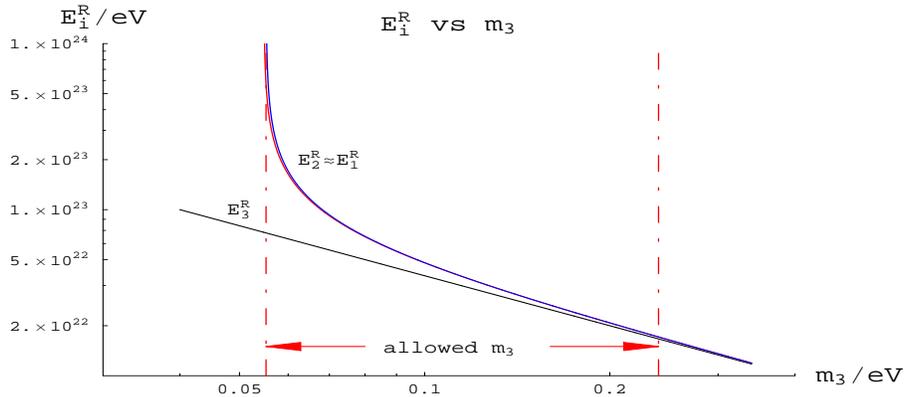}}}
 \caption{\small Resonant energies for different neutrino mass eigenstates
in the Z-burst model as a function of the largest neutrino eigenmass $m_3$. 
The $\Delta m^2$'s used here are the same as in Fig. 1.
}
 \label{fig:Zburst}
\end{figure}

The Z-burst mechanism \cite{Zburst} 
generates extreme-energy cosmic rays (EECR's)
by resonant annihilation of a EECR neutrino on the C$\nu$B neutrinos.
The resonant energy is 
\be{Zburst}
E^R = \frac{4\times 10^{21}\,{\rm eV}}{(m_\nu/{\rm eV})}\,.
\ee
The decay products of Z-bursts include on average two nucleons
and, from ten neutral pions, twenty photons.  
The decay multiplicity is $N\sim 30$.
The nucleons lose $f\sim 20\%$ of their energy for each 
$\lambda\sim 6$~Mpc traveled
in the CMB, so the average energy of a secondary nucleon arriving 
at Earth from distance $D$ is 
\be{nukenergy}
E_P\sim\frac{10^{21}{\rm eV}\times (0.8)^{D/6\,{\rm Mpc}}}{(m_\nu/0.1\,
{\rm eV})}\,.
\ee
The photons have shorter absorption lengths, except above $10^{21}$~eV,
and so are not expected to contribute much.
For a neutrino mass in the range of Eq.\ \rf{m3limits},
the mechanism is optimized:
a larger mass would move Z-burst secondaries down below the 
GZK energy $\sim {\rm few}\times 10^{19}$~eV where the ``background''
of normal EECR events is huge,
whereas a smaller mass would move the resonant energy beyond the reach 
of any realistic neutrino flux.
The Z-burst resonant energies as a function of the heaviest neutrino mass
$m_3$ are shown in Fig.\ \ref{fig:Zburst}.
Note that over most of the allowed $m_3$ range,
all three neutrinos contribute to annihilation 
with a resonant energy within a factor of two of each other.

In the simplest approximation, the spectrum of arriving nucleons is
\be{Zspectrum}
\frac{dN}{dE} \sim \frac{1}{D^2}\times\frac{dN}{dD}
\times\frac{dD}{dE}\;\;\propto\;\;E^{-1}
\ee
from sources uniformly distributed out to
\be{Dgzk}
D_{\rm GZK}\sim \lambda\,\frac{\ln\left(\frac{N\,E_{\rm GZK}}{E^R}\right)}{\ln (1-f)}\,,
\ee
 with a pileup at 
$E_{\rm GZK}$ resulting from all primaries originating 
beyond this distance.  
The $1/E$ spectrum extends from $E_{\rm GZK}$ out to the maximum 
nucleon energy $\sim E^R/30\sim 10^{21}(\frac{0.1{\rm eV}}{m_\nu})$~eV.
More realistic simulations including energy-loss processes,
cosmic expansion, and boosted Z-boson fragmentation functions 
give a softer spectrum, but a 
characteristic feature of the Z-burst mechanism remains 
that the super-GZK spectrum is considerably harder 
than the sub-GZK spectrum having power-law index -2.7.

What is not known is whether Nature has provided the large 
neutrino flux at $E^R$ to allow an appreciable event rate in 
future EECR detectors.
It is conceivable, although unlikely, that the flux is 
so large that present EECR events are initiated by Z-bursts.
A recent analysis \cite{FKR} of this possibility
gave a best fit with 
$m_{\nu}=0.26^{+0.20}_{-0.14}$~eV,
nicely consistent with the WMAP bound.
Another analysis \cite{Gelmini} fits the EECR spectrum down to the ankle
with Z-burst generated events and a neutrino mass of 0.07~eV,
again in accord with the WMAP bound.
The flux requirements for the Z-burst mechanism can be ameliorated
if there is an over-density of relic neutrinos, as would happen 
if (i) there was a significant chemical potential, 
or (ii) neutrinos were massive enough to cluster in ``local''
structures such as the Galactic SuperCluster.  Large chemical potentials
have been ruled out recently \cite{BellBeacom}, 
and this exclusion is confirmed by the WMAP data.  
Local clustering has been studied \cite{MaSingh},
with the conclusion being that a significant over density on the 
SuperCluster scale requires a neutrino mass in excess of 0.3~eV.
Such a mass is marginally allowed by the new WMAP/2dF limit.

\section{WMAP neutrino mass bound on \Rpv ~ SUSY} 

Supersymmetry without $R$-parity \cite{rpar} provides an elegant
mechanism for generating neutrino (Majorana) masses and mixings. In
these models, there are mainly two sources of neutrino mass
generation. In one scenario, products of trilinear $\lambda$ and/or
$\lambda'$ couplings generate a complete neutrino mass matrix through
one-loop self-energy graphs \cite{trilinear,recent}. 
In the other scenario,
the bilinear $R$-parity-violating terms induce sneutrino vacuum
expectation values (VEV's) allowing neutrinos to mix with the
neutralinos. 

The $L$-violating part of the \Rpv superpotential can be
written as 
\begin{equation}
    {\cal W}_{RPV}  =  {1\over 2}\lambda_{ijk} L_i L_j E^c_k
                        +  \lambda'_{ijk} L_i Q_j D^c_k + \mu_i L_i H_u, 
\label{superpot}
\end{equation}
where $i$, $j$ and $k$ are quark and lepton generation indices. In
Eq.~(\ref{superpot}), $L_i$ and $Q_i $ denote SU(2)-doublet lepton and
quark superfields, $E^c_i$ and $D^c_i$ are SU(2)-singlet charged
lepton and down-quark superfields, and $H_u$ is the Higgs superfield
responsible for the mass generation of the up-type quarks,
respectively. There are 9 $\lambda$-type (due to an antisymmetry in
the first two generation indices), 27 $\lambda'$-type and 3 $\mu_i$
couplings. Stringent upper limits exist on all these couplings from
different experiments \cite{review,faessler}.

We first consider the effects of the $\lambda'$ interactions. The
relevant part of the Lagrangian can be written as 
\begin{equation}
 - {\cal L}_{\lambda'} = \lambda'_{ijk} \left[\bar{d}_k P_L \nu_i
  \tilde{d}_{jL} + \bar{\nu}^c_i P_L d_j \tilde{d}^*_{kR}\right]
+ ~{\rm h.c.}   
\label{lagrangian}
\end{equation}
$P_L$ is the left-helicity projector.
Majorana mass terms for the left-handed neutrinos, given by ${\cal
L}_M = -\frac{1}{2} m_{\nu_{ii'}} \bar{\nu}_{Li} \nu^c_{Ri'} +~{\rm
h.c.}$, are generated at one loop. Figs. \ref{loop} show the
corresponding diagrams. The masses so induced are given by
\begin{equation} 
m_{\nu_{ii'}} \simeq {{N_c \lambda'_{ijk} \lambda'_{i'kj}}
\over{16\pi^2}} m_{d_j} m_{d_k}
\left[\frac{f(m^2_{d_j}/m^2_{\tilde{d}_k})} {m_{\tilde{d}_k}} +
\frac{f(m^2_{d_k}/m^2_{\tilde{d}_j})} {m_{\tilde{d}_j}}\right],
\label{mass}
\end{equation}     
where $f(x) = (x\ln x-x+1)/(x-1)^2$. Here, $m_{d_i}$ is the down quark
mass of the $i^{\rm th}$ generation inside the loop, $m_{\tilde{d}_i}$
is some kind of an
average of $\tilde{d}_{Li}$ and $\tilde{d}_{Ri}$ squark masses, and
$N_c = 3$ is the color factor. In deriving Eq.~(\ref{mass}), we
assumed that the left-right squark mixing terms in the soft part of
the Lagrangian are diagonal in their physical basis and are proportional
to the corresponding quark masses, {\em i.e.} $\Delta m^2_{\rm LR} (i)
= m_{d_i} m_{\tilde{d}_i}$. The small effect of quark mixing
is neglected in order not to
complicate the discussion unnecessarily. 

With $\lambda$-type interactions, one obtains exactly similar results
as in Eqs.~(\ref{lagrangian}) and (\ref{mass}). The quarks
and squarks in these equations will be replaced by the leptons and
sleptons of the corresponding generations. The color factor $N_c = 3$
would be replaced by $1$.
We do not explicitly write them down. 

For numerical purpose, we
have assumed the mass of whatever scalar is relevant to be 100 GeV
throughout, to be consistent with common practice and, in particular,
to compare with the old bounds. While for sleptons this sounds a
reasonable approximation, for squarks the present lower limit, even in
\Rpv ~scenarios, is around 250 GeV \cite{d0}. In any case, for different
squark masses one can easily derive the appropriate bounds by
straightforward scaling. It should be noted that the product couplings
under consideration contribute to charged lepton masses as well, but
with the present limits those contributions are too small to be of any
relevance. The resulting bounds are 
\ba {tri-bounds}
\lambda'_{i33}\lambda'_{i'33} < 3.6 \cdot 10^{-8},~~
\lambda'_{i32}\lambda'_{i'23} < 8.9 \cdot 10^{-7},~~
\lambda'_{i22}\lambda'_{i'22} < 2.2 \cdot 10^{-5},\nn \\
\lambda_{i33}\lambda_{i33} < 6.3 \cdot 10^{-7}, ~~
\lambda_{i32}\lambda_{i23} < 1.1 \cdot 10^{-5}, ~~
\lambda_{i22}\lambda_{i22} < 1.7 \cdot 10^{-4}.
\ea
There is one combination which receives a more stringent limit from 
$\mu e$ conversion in nuclei \cite{huitu}, namely
$\lambda'_{122}\lambda'_{222}<3.3 \cdot 10^{-7}$.
The chirality flips in 
Figs. \ref{loop} explain why with heavier fermions inside the loop
the bounds are tighter. For this reason, we have presented the bounds
only for $j, k= 2, 3$.  

Next we consider the bilinear $\mu_i$ terms. Such terms lead only to one
massive eigenstate as a result of tree level mixing between neutrinos
and neutralinos. The induced neutrino mass \cite{marta} is given by $m
\sim \mu_i^2/\mu$. Assuming the Higgsino mixing parameter $\mu = 100$
GeV, one obtains 
\be{bi-bounds} \mu_i/\mu < 1.5 \cdot 10^{-6}.  
\ee 
The bounds in Eqs.\ \rf{tri-bounds} and \rf{bi-bounds} obtained using
the recent WMAP bound are more stringent than the existing ones by one
order of magnitude, precisely to the extent that 
the WMAP data have improved the absolute neutrino mass bound.

We make a note in passing that even our improved bounds on trilinear
couplings do not invalidate the \Rpv~SUSY search strategies proposed by
the authors of \cite{barger}. Their suggestion is that at the Tevatron
collider the production and decay of sparticles would occur in R-parity
conserving modes except that the neutralino LSP would decay via
\Rpv~channel into multi-$b$ and missing energy final states constituting
the signal.

\section{Conclusions}
We have discussed implications of the WMAP neutrino bound on future
neutrino mass studies, including Tritium beta decay, neutrinoless double
beta-decay, and the Z-burst mechanism for EECR's.  We have shown that
the Tritium beta decay project KATRIN is unlikely to measure an absolute
neutrino mass, and that the WMAP bound in combination with the neutrino
oscillation data defines a predicted range for the double beta-decay
observable $m_{ee}$, which is accessible in the most ambitious proposed
experiments. The WMAP bound also impacts the Z-burst mechanism for
cosmic rays above the GZK cutoff. It constrains local over-densities, but
it also limits the resonant energy to a favorable range.

Turning to model building, WMAP constrains theories with $\Delta L=1$
lepton number violation, since in these theories Majorana neutrino
masses are generated radiatively.  Taking \Rpv~SUSY as our example, we
have derived the upper limits on many individual and product couplings
of the $\lambda$- and $\lambda'$-types, and also the bilinear $\mu_i$
terms, from their contribution to neutrino masses.  Using the recent
WMAP bound the limits have been improved by an order of magnitude.
Finally, we remark that the new WMAP bound on neutrino mass coincides
nicely with the one arising from the requirement of successful
baryogenesis in the context of the neutrino see-saw model
\cite{baryogen}.

\section*{Note Added}
The new mass upper-limit expressed in Eqs.\ \rf{mapbound}, \rf{mapmass},
and \rf{m3limits} depends on priors.
In particular, the bound depends on the inclusion of Ly$\alpha$ data 
to estimate the power spectrum of intervening hydrogen clouds.
Given the complexity of the Ly$\alpha$ analysis, 
some have questioned the reliability of this data set.  
Without the Ly$\alpha$ prior,
the WMAP mass limit is relaxed to 
${\cal O}(1)$~eV \cite{noLya}.  
Accordingly, the upper dot-dash lines in Figs.\ \ref{fig:mapbound}
and \ref{fig:Zburst}, 
and the upper bounds in Eqs.\ \rf{degenmass}, \rf{tri-bounds},
and \rf{bi-bounds}, would also relax by $\sim 40\%$.

\section*{Acknowledgments}
We thank Nicole Bell and Bill Louis for helpful remarks.  GB
acknowledges hospitality of LPT, Univ. de Paris XI, Orsay, where a part
of the work has been done.  This research has been supported by (i)
U. S. Department of Energy grant number DE-FG05-85ER40226, (ii) the
Bundesministerium f\"ur Bildung und Forschung (BMBF, Bonn Germany) under
contract number 05HT1WWA2 (for HP), and (iii) in part, the Department of
Science and Technology, India, project number SP/S2/K-10/2001 (for GB).


\begin{figure}
\begin{center}
\begin{picture}(500,600)(0,0)
\SetWidth{2.0}
\ArrowLine(60,450)(140,450)
\ArrowArc(220,450)(80,0,90)
\ArrowArc(220,450)(80,90,180)
\DashLine(140,450)(220,450){3}
\DashLine(220,450)(300,450){3}
\ArrowLine(380,450)(300,450)
\Text(220,452)[]{$\Huge{\boldmath{{\bullet}}}$}
\Text(220,532)[]{$\Huge{\boldmath{{\otimes}}}$}
\Text(100,470)[]{$\Large{\boldmath{\nu_{_{iL}}}}$}
\Text(340,470)[]{$\Large{\boldmath{\nu_{_{i'L}}}}$}
\Text(155,525)[]{$\Large{\boldmath{d_{jL}}}$}
\Text(295,525)[]{$\Large{\boldmath{d_{jR}}}$}
\Text(180,430)[]{$\Large{\boldmath{\tilde{d}_{kR}}}$}
\Text(260,430)[]{$\Large{\boldmath{\tilde{d}_{kL}}}$}
\Text(140,430)[]{$\Large{\boldmath{\lambda'_{ijk}}}$}
\Text(300,430)[]{$\Large{\boldmath{\lambda'_{i'kj}}}$}
\Text(220,380)[]{\LARGE{\bf{(a)}}}

\ArrowLine(60,150)(140,150)
\ArrowArcn(220,150)(80,90,0)
\ArrowArcn(220,150)(80,180,90)
\DashLine(140,150)(220,150){3}
\DashLine(220,150)(300,150){3}
\ArrowLine(380,150)(300,150)
\Text(220,151)[]{$\Huge{\boldmath{{\bullet}}}$}
\Text(220,231)[]{$\Huge{\boldmath{{\otimes}}}$}
\Text(100,170)[]{$\Large{\boldmath{\nu_{_{iL}}}}$}
\Text(340,170)[]{$\Large{\boldmath{\nu_{_{i'L}}}}$}
\Text(155,225)[]{$\Large{\boldmath{d_{kR}}}$}
\Text(295,225)[]{$\Large{\boldmath{d_{kL}}}$}
\Text(180,130)[]{$\Large{\boldmath{\tilde{d}_{jL}}}$}
\Text(260,130)[]{$\Large{\boldmath{\tilde{d}_{jR}}}$}
\Text(140,130)[]{$\Large{\boldmath{\lambda'_{ijk}}}$}
\Text(300,130)[]{$\Large{\boldmath{\lambda'_{i'kj}}}$}
\Text(220,80)[]{\LARGE{\bf{(b)}}}
\end{picture}
\end{center}
\caption{\small The $\lambda'$-induced one loop diagrams
contributing to Majorana masses for the neutrinos. The $\lambda$-induced
diagrams are analogue with sleptons propagating in the loop.
\label{loop}}
\end{figure}

\begin{thebibliography}{99}

\bibitem{Spergel:2003cb}
D.~N.~Spergel {\it et al.},
astro-ph/0302209.

\bibitem{cbi-acbar}
T.~J.~Pearson {\it et al.},
astro-ph/0205388;
C.~l.~Kuo {\it et al.}  [ACBAR collaboration],
astro-ph/0212289.



\bibitem{2dF}
M. Colles {\it et al.}, 2001, MNRAS, 328, 1039; 
O. Elgaroy et al., Phys. Rev. Lett. 89, 061301 (2002).

\bibitem{fournu} A. Pierce and H. Murayama, hep-ph/0302131.

\bibitem{giunti}
C. Giunti, hep-ph/0302173.

\bibitem{abazajian} 
K. N. Abazajian, Astropart.\ Phys., to appear [astro-ph/0205238].

\bibitem{BBN}
S. Hannestad, astro-ph/0302340.

\bibitem{DiBariAnd} P. Di Bari, Phys. Rev.\ {\bf D65}, 043509 (2002), 
and addendum in arXiv:astro-ph/0302433;
also, Nicole Bell, private communication.

\bibitem{lepasym}
R.~Foot and R.~R.~Volkas,
Phys.\ Rev.\ Lett.\  {\bf 75}, 4350 (1995).


\bibitem{PSW}
For some details on the viability of the sterile neutrino, see
H. P\"as, L. Song, T.J. Weiler, hep-ph/0209373 and references therein.

\bibitem{PW} H. P\"{a}s and T. J. Weiler, Phys.\ Rev.\ D63, 113015 (2001);
for more details, see S. M. Bilenky, C. Giunti, J. A. Grifols, and E. 
Masso, hep-ph/0211462.

\bibitem{gautam}
G.~Bhattacharyya, H.~V.~Klapdor-Kleingrothaus and H.~P\"as,
Phys.\ Lett.\ B 463, 77 (1999).

\bibitem{katrin}
C. Weinheimer, Nucl. Phys. Proc. Suppl. 91, 273 (2001);
A. Osipowicz {\it et al.}, (KATRIN Collab.), hep-ex/0109033;
C. Weinheimer, private communication.

\bibitem{Fornengo:2001pm}
N.~Fornengo, M.~Maltoni, R.~T.~Bayo and J.~W.~Valle,
Phys.\ Rev.\ D 65, 013010 (2002).

\bibitem{solar}
M.~Maltoni, T.~Schwetz and J.~W.~Valle,
hep-ph/0212129;
A. Bandyopadhyay {\em et al.}, hep-ph/0212146;
J.~N.~Bahcall, M.~C.~Gonzalez-Garcia and C.~Pena-Garay,
hep-ph/0212147;
V. Barger and D. Marfatia, Phys.\ Lett.\ B555, 144 (2003);
P.C. de Holanda and A.Y. Smirnov, hep-ph/0212270; H. Nunokawa {\em et
al.}, hep-ph/0212202; P. Aliani {\em et al.}, hep-ph/0212212. 

\bibitem{chooz} CHOOZ collaboration, M. Appolonio {\em et al.}, Phys.
Lett. B 466, 415 (1999).



\bibitem{kps}
H.V. Klapdor-Kleingrothaus, H. P\"as, A.Yu. Smirnov,
Phys. Rev. D~63 (2001) 073005;
V. Barger, S.L. Glashow, D. Marfatia, and K. Whisnant,
Phys.\ Lett.\ B532, 15 (2002).

\bibitem{GENIUS}
H.~V.~Klapdor-Kleingrothaus, J.~Hellmig and M.~Hirsch,
J.\ Phys.\ G 24 (1998) 483;
H. V. Klapdor-Kleingrothaus, L. Baudis, G. Heusser, B. Majorovits, 
H. P\"as, hep-ph/9910205.

\bibitem{elliot}
S.R. Elliott, P. Vogel, hep-ph/0202264, Annu. Rev. Nucl. Part. Sci., 
vol.52.

\bibitem{evi}
H.V. Klapdor-Kleingrothaus, A. Dietz, H.L. Harney, I.V. Krivosheina,
Mod. Phys. Lett.  A 16 (2001) 2409. 
Compare, however, criticisms in
C.E. Aalseth {\it et al.}, hep-ex/0202018 and
F. Feruglio, A. Strumia, F. Vissani, hep-ph/0201291;
H.L. Harney, hep-ph/0205293; with the rebuttal in
H.V. Klapdor-Kleingrothaus, hep-ph/0205228; 
ibid.\ Foundations of Physics 32, 1181 (2002).

\bibitem{extra}
G. Bhattacharyya, H.V. Klapdor-Kleingrothaus, H. P\"as, A. Pilaftsis,
hep-ph/0212169.

\bibitem{znbbrev}
H.V. Klapdor--Kleingrothaus, H. P\"as, hep-ph/0002109,
in: Proc. of the {\it International Workshop on Particle Physics and
the Early Universe}, September 27 - October 2, 1999, 
Trieste, Italy.


\bibitem{Zburst} 
T.J. Weiler, Phys. Rev. Lett. 
49 (1982) 234;
Astroparticle Phys. 11 (1999) 303;
D. Fargion, B. Mele, A. Salis, Astrophys. J. 517 725 (1999);
T.J. Weiler, 
in ``Beyond the Desert 99, Ringberg Castle'', Tegernsee, Germany, 
June 6-12, 1999 [hep-ph/9910316], 
E. Roulet, Phys. Rev. D 47 5247, (1993); 
S. Yoshida, H-y. Dai, C.C.H. Jui, 
P. Sommers, Astrophys. J. 479, 547 (1997). 


\bibitem{FKR} Z. Fodor, S. D. Katz, and A. Ringwald, JHEP 0206, 046 (2002).

\bibitem{Gelmini} G. Gelmini and G. Varieschi, hep-ph/0201273.

\bibitem{BellBeacom}
A.D. Dolgov et al., Nucl.\ Phys.\ B632, 363 (2002);
%
Y. Y. Wong, Phys.\ Rev.\ D66, 025015 (2002); 
%
K. N. Abazajian, J. F. Beacom, and N. F. Bell, 
Phys.\ Rev.\ D66, 013008 (2002). 

\bibitem{MaSingh} S. Singh and C.-P. Ma, Phys.\ Rev.\ D67, 023506 (2003).

\bibitem{rpar} G. Farrar, P. Fayet, Phys. Lett. 76B, 575 (1978); 
S. Weinberg, Phys. Rev. D 26, 287 (1982);
N. Sakai, T. Yanagida, Nucl. Phys. B 197, 533 (1982);
C. Aulakh, R. Mohapatra, Phys. Lett. 119B, 136 (1982).

\bibitem{trilinear} S. Dimopoulos, L. Hall, Phys. Lett. B207, 210
(1987); R. Godbole, P. Roy, X. Tata, Nucl. Phys. B 401, 67 (1993). 

\bibitem{recent} M. Drees, S. Pakvasa, X. Tata
and T. ter Veldhuis, Phys. Rev. D 57, 5335 (1998); S. Rakshit,
G. Bhattacharyya, A. Raychaudhuri, Phys. Rev. D 59, 091701 (1999);
R. Adhikari, G. Omanovic, Phys. Rev. D 59, 073003 (1999); O. Kong,
Mod. Phys. Lett. A 14, 903 (1999); E.J. Chun, S.K. Kang, C.W. Kim,
U.W. Lee, Nucl. Phys. B 544, 89 (1999); K. Choi, K. Hwang,
E.J. Chun, Phys. Rev. D 60, 031301 (1999); A. Joshipura, S. Vempati,
Phys. Rev. D 60, 111303 (1999).

\bibitem{review} For reviews, see
G. Bhattacharyya, hep-ph/9709395, Proc. BEYOND 97, Castle
Ringberg, Germany, June 1997, eds.~H.V. Klapdor-Kleingrothaus,
H. P\"{a}s, IOP, Bristol, 1998 (and its update in the
Proc. of the Int.~WEIN98 Conf., Santa Fe, USA, June 1998);
hep-ph/9608415;  
H. Dreiner, hep-ph/9707435;
R. Barbier {\it et al.}, hep-ph/9810232;
P. Roy, hep-ph/9712520. See also, V. Barger, G.F. Giudice and T. Han,
Phys. Rev. D 40, 2987 (1989).  

\bibitem{faessler} V. Bednyakov, A. Faessler, S. Kovalenko,
Phys. Lett. B 442, 203 (1998). 


\bibitem{d0} B. Abott {\em et al.}, D0 Collaboration, 
Phys. Rev. Lett. 83, 4476 (1999). 

\bibitem{huitu} A. Faessler, T.S. Kosmas, S. Kovalenko,
J.D. Vergados, hep-ph/9904335; K. Huitu, J. Maalampi, M. Raidal,
A. Santamaria, Phys. Lett. B 430, 355 (1998).

\bibitem{marta} S. Davidson and M. Losada, Phys. Rev. D 65, 075025 (2002).

\bibitem{barger} V. Barger, T. Han, S. Hesselbach and D. Marfatia, 
Phys. Lett. B 538, 346 (2002).

\bibitem{baryogen} 
W. Buchm\"uller, P. Di Bari, and M. Pl\"umacher,
Phys.\ Lett.\ B547, 128 (2002);
hep-ph/0302092.

\bibitem{noLya} S. Hannestad, arXiv:astro-ph/0303076;
O. Elgaroy and O. Lahav, arXiv:astro-ph/0303089.

\end{thebibliography}
\end{document}